\documentstyle[12pt,fleqn]{article}
\newcommand\beq{\begin{equation}}
\newcommand\eeq{\end{equation}}

\newcommand\half{{1\over 2}}
\newcommand\twothirds{{2\over 3}}

\newcommand\dvvec{\stackrel{\leftrightarrow}{\partial_\mu}}

\newcommand\be{\begin{eqnarray}}
\newcommand\ee{\end{eqnarray}}
\newcommand{\equa}[1]{(\ref{#1})}

\newcommand{\cL}{{\cal L}}

\newcommand{\del}{\partial}

\newcommand{\prd}[3] {{Phys.\,Rev.} {\bf D#1} (19#3) #2}
\newcommand{\prold}[3] {{Phys.\,Rev.} {\bf #1} (19#3) #2}
\newcommand{\plb}[3] {{Phys.\,Lett.} {\bf #1 B} (19#3) #2}
\newcommand{\prl}[3] {{Phys.\,Rev.\,Lett.} {\bf #1} (19#3) #2}

\newcommand{\npb}[3] {{Nucl.\,Phys.} {\bf B#1} (19#3) #2}

\begin{document}
\title{The Cheshire Cat Bag Model:
Color Anomaly and $\eta'$ Properties\thanks{Talk presented at the
International Nuclear Physics Conference,  Wiesbaden (Germany), July 26 -
Aug.~1, 1992}}
\author{A. Wirzba\\Nordita
\\Blegdamsvej 17\\
DK-2100 Copenhagen {\O}\\Denmark}
\date{}
\maketitle
\vfill
\begin{abstract}
We show that color can leak from a QCD bag if we allow for pseudoscalar
isoscalar singlet ($\eta'$) coupling at the surface. To enforce total
confinement of color an additional boundary term is suggested. New
relations between the $\eta'$ mass and decay constant and the QCD gluon
condensates are derived and compared with the empirical parameters.\\ \\
\end{abstract}
\vfill
\begin{flushleft}
Nordita - 92/68 N\\
hep-ph/yymmnn
\end{flushleft}
\newpage
\noindent{\bf 1. INTRODUCTION}\\

The bag model was originally introduced in order to have color
confinement by hand \cite{MIT}: the bag exterior acts as a perfect color
dielectric, while the color degrees of freedom are confined to the bag
interior. The latter has been achieved by forbidding color to flow from
the bag interior to the exterior.  This means that, at the surface of
the bag, the gluons are subject to the following confining boundary
equations
\beq
 \hat n \cdot {\vec E}^a =0, \quad \hat n \times {\vec B}^a=0,
  \label{MITG}
\eeq
where $\hat n$ is the outward normal to the bag surface. The color
electric fields $E^a_\mu$ have to point along the surface of the bag
($a$ is the color index), while the color magnetic fields $B_\mu^a$ have
to be orthogonal to the surface.  For the quarks, the confining boundary
equation states that the flux of the  isoscalar  current through the bag
surface is forbidden: $ n_\mu {\bar q} \gamma^\mu q =0$, where $n^\mu=
(0,\hat n)$ is the space-like 4-vector generalization of  the normal
vector at the bag surface. In comparison to this  traditional picture of
the  bag model, the bag from the ``Cheshire cat" point of view serves
just as a formal separation of space-time in a {\em bag region} where
the underlying QCD still acts, an {\em exterior region} where QCD is
``integrated out" (practically meaning ``replaced") by an effective
action  formulated in hadronic degrees of freedom, and a {\em boundary}
where both regimes are matched so generally that it doesn't matter where
the bag walls are located \cite{cheshire,N87,NW87}. In other words, the
bag itself has  no physical significance,  only its formalism is used.
Therefore the name ``Cheshire cat bag" as  inspired by the vanishing of
the Cheshire cat in the Lewis Caroll's  tale {\em Alice in Wonderland}
\cite{alice}. Of course it should be clear that in $3+1$ dimensions the
Cheshire cat bag can only work approximately in practice, since we know
from {\it e.g.} large $N_c$ arguments that an effective action trying to
replace QCD must eventually involve infinitely many local terms.  In the
Cheshire cat formulation of the bag model it is obvious that the
confining conditions for gluons and quarks will be not satisfied in
general, since the bag wall can be moved at will. \\ \\

\noindent{\bf 2. TOY MODEL IN 1+1 DIMENSIONS}\\

As mentioned, we can at most expect an approximate formulation of the
Cheshire cat bag model in the physical $3+1$ space-time, since we
have only an approximate bosonization of QCD at our disposal,
as inspired by large $N_c$ arguments.
In the $1+1$ dimensional toy world, however, fermion and boson
formulations are exactly equivalent. Because of this, we can expect that
the $1+1$ dimensional analogon of the $3+1$ Cheshire cat bag will work
exactly \cite{N87,NW87}. For that purpose, let us start out with
the simplest case:
massless fermions in the bag interior, massless bosons in the exterior,
and the usual hybrid bag model coupling  \cite{cbag} between the
fermions and bosons at the boundary in order to guarantee chiral
invariance. The bag lagrangian reads therefore
\beq
 \cL_{\rm bag} = \left ( i \bar \psi \half\gamma^\mu \dvvec \psi \right)
                         \Theta_{\rm in}
                  -\half \bar \psi e^{i \gamma_5 \phi/f} \psi \Delta_S
                +\left (\half \del_\mu \phi \del^\mu \phi \right)
                         \Theta_{\rm out}
 \label{lone}
\eeq
where $\psi$ is the fermion field, $\phi$ the (pseudo-)scalar boson
field, $\Theta_{\rm in/out}$ are step-functions with support on the bag
interior/exterior and $\Delta_S$ is the surface delta-function. We know
that massless fermions can be exactly bosonized into massless bosons --
both theories are equivalent. Thus we can ask the question what will
happen to a e.g.\ right-moving (right-handed) fermion wave packet
prepared in the bag interior,  when it
hits the bag wall \cite{NW87}.
The underlying bosonization arguments would tell that
the fermion wave packet should pass the bag wall in an undisturbed
fashion. On the other hand, the linear quark boundary condition
\beq
   -i n_\mu \gamma^\mu \psi (t,x_{\rm bw})
 = e^{i \gamma_5\phi(x_{\rm bw})/f} \psi(t,x_{\rm bw})
\eeq
in the Weyl representation
for the $\gamma$-matrices
implies that in the usual static bag approximation
the right moving (right handed)
quark wave packet $\psi_R$ is -- modulo a constant phase shift --
completely reflected into a left moving
(left handed) one $\psi_L$ at the bag surface:
\beq
   i \psi_L (t+x)= e^{i \phi/f} \psi_R(t+x-2 x_{\rm bw}).
\eeq
This paradox can be resolved, if the boundary condition is generalized
to involve time-dependent $\phi$-fields, $\phi(t,x_{\rm bw})$.
In this case the $\phi$-field can be Taylor-expanded  around the
collision time $t_0$ of the center of the wave packet to linear order
such that the reflected particle gets an extra time-dependent (an via
translation also space-dependent) phase factor $\exp(i (t+x)\dot \phi/f)$.
This means that the wave packet is not only reflected at the bag wall,
but also acquires an extra momentum $\Delta p = \dot \phi /f$ and an
extra energy
kick $\Delta E = -\dot \phi/f$ via the time-dependent boundary.
At the same time  the temporal shift in $\phi$ causes
a boson excitation
to be sent to the right.  Using phase space arguments it can be shown that
when exactly one fermion state  is ``drowned" in the Dirac
sea ($\Delta Q=1$), the boson
excitation carries exactly one winding number, $\Delta \phi = 2\pi
f$:
\beq
   \Delta Q = \frac{1}{h} \Delta p \Delta x = \frac{\dot\phi}{2 \pi f} c
    \Delta t  \equiv \dot Q \Delta t,
    \label{Banom}
\eeq
with Planck's constant $h=2\pi$ and the velocity of light $c=1$.
Thus the paradox is solved by ``drowning" the reflected fermion wave
packet in the Dirac sea and by simultaneously exciting a solitonic boson
configuration which carries now (via bosonization) the information
that originally a right-moving fermion wave packet was created.
{}From eq.\equa{Banom}\ one
reads off the following fermion number anomaly,
$\dot Q = \dot\phi/(2\pi f)$ valid at the bag wall.
Let us now extend the model by coupling a U(1) gauge field
$A_\mu(x,t)$ to the fermions in the bag interior, $e\bar \psi \gamma^\mu
A_\mu \psi$. This so-called Schwinger model can still be exactly
bosonized into a -- now -- massive boson field in case the boson mass in
the term $\half m^2\phi^2$ is fine-tuned to $m=e/\sqrt{\pi}$. Is the
usual hybrid bag coupling (see eq.\equa{lone}) still sufficient to
guarantee a Cheshire cat bag situation? The answer is no: Since the
fermions carry not only fermion number charge, but now also electric
charge, the fermion number anomaly \equa{Banom}\ induces an anomaly in
the electric charge $ e\int^t dt' \dot \phi/(2\pi f)$. This
non-conserved electric charge must be compensated by a counter term
which can only act at the bag surface, since the boson field itself is
charge-neutral and since the fermion number anomaly (and thus the charge
anomaly, too) acts at the boundary: ${\cal L}_{\rm CT} = e A_0
\phi/(2\pi f) \Delta_S$. Covariantly, the counter lagrangian reads
\beq
  {\cal L}_{\rm CT} = -\frac{e}{2\pi} \epsilon^{\mu \nu} n_\nu A_\mu
   \frac{\phi}{f} \Delta_S,
\eeq
where $\epsilon^{\mu\nu}$ is the anti-symmetric tensor in 1+1 dimensions.
Note that
the counter term is not gauge invariant in order to cancel the non-gauge
invariance resulting from the charge anomaly of the fermions.
As a consequence of the new counter term the electric field $E$
satisfies the following generalized boundary equation
\beq
  E = -\frac{e}{2\pi} \frac{\phi}{f}.
  \label{newone}
\eeq
at the bag surface instead of the 1+1 dimensional analog of
the M.I.T.\ boundary equations \equa{MITG},
$E = 0$.
\\ \\

\noindent{\bf 3. GENERALIZATION TO 3+1 DIMENSIONS}\\

We will here only present a heuristic argument for the color anomaly. An
exact argument based on the multiple reflection method \cite{HJ83}
and completely analogous to the fractional baryon number calculation
of Goldstone and Jaffe \cite{GJ83} can be found in ref.\cite{coloranom}.

Let as assume that there is a non-zero color magnetic field $\vec B^a$
present in the neighborhood of the bag, given in some fixed gauge which
will be the same as the gauge of the color charge in the color anomaly.
Because of
the usual M.I.T.\ boundary condition for the gluons \equa{MITG}, the
color magnetic fields point perpendicular to the bag surface and the
quarks close to the bag surface
grouped in the corresponding Landau levels will do the same.
There are two types of Landau levels, the lowest ones have a linear
dispersion, while the rest have a parabolic dispersion. Whereas the
latter are passive for the argument to be presented, the former behave
exactly as the fermion levels in the  1+1 dimensional  bag in case we
couple the quarks  at the bag surface to the pseudo-scalar isoscalar 3+1
dimensional $\eta'$ field:
$  -\half \bar q \exp(i\gamma_5 \eta'/f) q \Delta_S$.
If the $\eta'$ field at the bag surface is time-dependent,
the ``reflected" quarks moving
in the linear Landau levels get again a ``phase kick" in energy and
momentum and there will be an anomaly in the quark number. The latter
will induce a color anomaly, since the quarks carry color charge. Its
value can be found by transcribing the integrated fermion number
anomaly,  $e \phi/(2\pi f)$, from 1+1 dimensions to the 3+1 dimensional
situation, $\tilde g(s) \eta'/(2\pi f)$  and multiplying it with the
number of Landau states $N_F \tilde g(s) \hat n \cdot\vec B^a /(2 \pi)$
per area along the bag wall. $N_F$ is the effective number of light
flavors and $\tilde g(s)$ is the quark-gluon coupling constant ($\tilde
g(s)^2 =\half g(s)^2$ in terms of the gluon coupling constant which is
defined for a given scale $s$ associated with the size of the bag
surface). The result is:
\beq
 \Delta Q^a = N_F \frac{g(s)^2}{8 \pi^2} \oint_{\rm Area} d\Sigma_\beta\,
  \vec B^a(\beta) \cdot \hat n_\beta
  \frac{\eta'(\beta)}{f}
\eeq
with $d \Sigma_\beta$ denoting the area element at the surface point
$\beta$. Thus contrary to the common belief the chiral boundary $\{i
n_\mu \gamma^\mu +\exp(i \gamma_5 \eta'/f)\}q(\beta) =0$ ensures the
conservation of color charge only at the classical level, but not at
the quantum level. In order to restore gauge invariance at the quantum
level a non-gauge invariant counter term acting at the bag surface has
to be added. Here is
the final result of ref.\cite{coloranom} which is formulated in terms of
the Chern-Simons current $K^\mu_5 = \epsilon^{\mu\nu\alpha\beta}
(A^a_\nu G^a_{\alpha\beta} -\twothirds f^{abc} A^a_\nu A^b_\alpha
A^c_\beta )$
and is the non-abelian
generalization corresponding to the above derived result:
\beq
 {\cal L}_{\rm CT} = \frac{g(s)^2}{16\pi^2}
            \oint_{\Sigma} d\Sigma_\beta K^\mu_5
         n_\mu \frac{\eta'}{f}
  \label{CTthree}
\eeq
As consequence of the new counter term the restricted M.I.T.\ boundary
equations for the gluons \equa{MITG}\ are now replaced by
\be
     \hat n \cdot  \vec E^a &=& -\frac{N_F g^2(s)}{8\pi^2 f}
                               \hat n \cdot\vec B^a \eta'
               \label{newE}\\
     \hat n \times \vec B^a &=& \frac{N_F g^2(s)}{8\pi^2 f}
                               \hat n \times\vec E^a \eta'.
                \label{newB}
\ee
Note that for $\eta' = 0$ the old M.I.T.\ boundary terms are recovered.
\\ \\

\noindent{\bf 4. CHESHIRE CAT DERIVATION OF THE $\eta'$ MASS}\\

In the following we will combine the new derived boundary terms
\equa{newE}\ and \equa{newB}\
with the Cheshire cat principle in order to derive
the $\eta'$ mass. For this purpose the bag is used as a  {\em test bag}
which replaces in a small spatial volume of variable size $V$ the
profile of a static $\eta'$ solution which still acts in the remainder
of space. This should be done in such a way that after tuning the
parameters of the $\eta'$ field (here the mass) it doesn't matter
whether the bag is inserted or not. Integrating the (linearized) static
equation of  motion of the $\eta'$ field over the volume $V$, we get the
following set of relations
\beq
 \int_V  m^2 \eta' = \int_V \nabla^2 \eta' = \oint_{\partial V} d \vec
  \Sigma \cdot \vec \nabla \eta' = -\frac{1}{2f}\oint_{\partial V}d \vec
   \Sigma \cdot \vec j^5_{\rm bos}
  \label{bosside}
\eeq
where in the last step the bosonic expression for the axial current has
been inserted. Because of continuity in the axial current at the bag
surface, the fluxes of the boson and quark axial current are the same
and  we have
\beq
 {\rm eq.}\equa{bosside} = -\frac{1}{2f}\oint_{\partial V}d \vec
   \Sigma \cdot \vec j^5_{\rm quark}= -\frac{1}{2f} \int_V {\rm Anomaly}
\eeq
where the fact has been used that the axial current is anomalous,
$\del_\mu j^{5\mu}= {\rm Anomaly}$. The
term $\int_V \dot q^5/(2f)$ which would normally have appeared on
the right hand side of the last equation has been dropped,
since a time-variation in the axial charge $\int_V q^5$ would be in
contradiction to the
{\em static} ansatz. Since the location and
the size of the volume $V$ can be chosen at will, we finally have the
local operator identity
\beq
 m^2 \eta' = -\frac{1}{2 f} {\rm Anomaly}
\eeq
involving the usual chiral anomaly. Up to this point the derivation is
completely general, valid for 1+1 (where $\eta'$ should be replaced
by $\phi$ of
course) as well as for 3+1 dimensions. If we (a) insert for the chiral
anomaly the 1+1 dimensional expression $(e/\pi)E$, (b) relate via the
new boundary term \equa{newone}\ $E$ locally to  $E= - e \phi/(2\pi f)$
and (c) use the fact that in the 1+1 case the decay constant $f$ is just
a $c$-number,  $f=1/\sqrt{4\pi}$ (this can again be derived via the
Cheshire cat principle see refs.\cite{cheshire,N87}), we end up with the
relation $m^2\phi=(e^2/\pi) \phi$. Thus the Cheshire cat derivation in
1+1 dimensions gives the exact Schwinger model result $m^2=e^2/\pi$
without ever making explicit use of the underlying bosonization.  In the
3+1 case the chiral anomaly has the form $N_F (g^2/4 \pi^2) \vec
E^a\cdot \vec B^a$. Under the M.I.T.\ boundary terms \equa{MITG}\ the
scalar product $\vec E^a\cdot \vec B^a$  would  normally be zero; under
the new boundary terms \equa{newE}\ and \equa{newB}, however, this
product can be re-expressed by the gluon condensate and the $\eta'$
field: $N_F (g^2/16\pi^2)\langle G^2 \rangle \eta'/f$, see
ref.\cite{tale} for more details where also the final result for the
$\eta'$ mass in the 3+1 dimensional world is listed:
\beq
 f^2 m^2 = N_F \frac{g(s)^2}{8 \pi^2} \langle 0| \frac{g(\mu)^2}{8\pi^2}
 G(\mu)^2 |0\rangle.
\eeq
Inserting in this relation (which is analogous to the
Gell-Mann-Oakes-Renner relations for the octet Goldstone bosons
\cite{GOR}) the
value $\langle 0| (\alpha(\mu)/\pi) G(\mu)^2 |0\rangle \sim (330\, {\rm
MeV})^4$ borrowed from the QCD sum rule calculations \cite{sumrules} and
for $g(s)^2$
values for reasonable bag scales (see ref.\cite{tale}) one ends up with
$\eta'$ masses between 0.33~GeV and 1.6~GeV compared to the empirical
value of 0.96~GeV. In this calculation the $\eta'$ decay constant was in
turn derived from the Cheshire cat principle and had
the reasonable value $f_{\eta'} \equiv \sqrt{2/N_F} f = 100\,{\rm MeV}$.
\\ \\

\noindent{\bf 5. CONCLUSION}\\

There is necessarily a color anomaly in any hybrid bag model with a
chiral coupling to a pseudoscalar isoscalar ($\eta'$) field at the bag
surface. Consequently, a gauge-dependent counter term is required in
order to restore the overall gauge invariance. The counter term induces
changes in the gluon boundary equations, see \equa{newE}\ and
\equa{newB}\  which relate the M.I.T.~forbidden fields at the bag
surface to the M.I.T.~allowed ones times the $\eta'$ field. The old
M.I.T.~boundary equations \equa{MITG}\ are recovered for a vanishing
$\eta'$ field. As a consequence non-zero matrix-elements $\langle0| \vec
E^a \cdot \vec B^a| \eta'\rangle$ can easily exist in the bag under the
new boundary terms  in contrast to usual bag calculations. Furthermore,
even for a spherical bag the color electric field can now point
radially. Finally, using the new boundary terms combined with the
Cheshire cat principle the $\eta'$ mass, and -- as shown in
ref.\cite{tale} -- decay constant $f_{\eta'}$  and 4-point vertex can be
derived in reasonable agreement with the empirical information in 3+1
dimensions where the Cheshire cat principle is expected to work
approximately and in exact agreement with the bosonization results in
the 1+1 dimensional case. \\

Most of this presentation is based on the work of
refs.\cite{NW87,coloranom,tale}. The author would like to express his
gratitude to the
co-authors of these papers,  Holger Bech Nielsen,
Mannque Rho and Ismail Zahed.


\begin{thebibliography}{99}
%
\bibitem{MIT} A. Chodos, R.L. Jaffe, K. Johnson, C.B. Thorn and V.
Weisskopf,
\prd{9}{3471}{74}.
\\[-8mm]
%
\bibitem{cheshire} S. Nadkarni, H.B. Nielsen and I. Zahed,
\npb{253}{308}{85};\\
S. Nadkarni and H.B. Nielsen, \npb{263}{1}{86};\\
S. Nadkarni and I. Zahed, \npb{263}{23}{86};\\
R.J. Perry and M. Rho, \prd{34}{1169}{86}.
\\[-8mm]
%
%
\bibitem{N87} H.B. Nielsen, in ``Workshop on Skyrmions and Anomalies'',
Krak{\'{o}}w, Poland 20-24 February, 1987, World Scientific, Singapore
(1987), pp. 277.
\\[-8mm]
%
\bibitem{NW87} H.B. Nielsen and A. Wirzba, in
``The Elementary Structure of Matter,''
Workshop Proceedings (Les Houches, March 24 -- April 2, 1987),
Springer Proceedings in Physics {\bf 26}, Springer, Berlin (1988).
\\[-8mm]
%
\bibitem{alice} Lewis Caroll, ``Alice's Adventures in Wonderland",
Macmillan Publ., London, 1865.
\\[-8mm]
%
\bibitem{cbag} A. Chodos and C.B. Thorn, \prd{12}{2733}{75};\\
G.E. Brown and M. Rho, \plb{82}{177}{79}.
\\[-8mm]
%
\bibitem{coloranom} H.B. Nielsen, M. Rho, A. Wirzba and I. Zahed,
\plb{269}{389}{91}.
\\[-8mm]
%
\bibitem{HJ83} T.H. Hansson and R.L. Jaffe, \prd{28}{882}{83}.
\\[-8mm]
%
\bibitem{GJ83} J. Goldstone and R.L. Jaffe, \prl{51}{1518}{83}.
\\[-8mm]
%
\bibitem{tale} H.B. Nielsen, M. Rho, A. Wirzba and I. Zahed,
\plb{281}{345}{92}.
\\[-8mm]
%
\bibitem{GOR} M. Gell-Mann, R.J. Oakes and B. Renner,
\prold{175}{2195}{68}.
\\[-8mm]
%
\bibitem{sumrules} M.A. Shifman, A.I. Vainshtein and V.I. Zakharov,
\npb{147}{385, 448, 519}{79}; L. Reinders, Acta Physica Polnica
{\bf B15} (1989) 329.
%
\end{thebibliography}
\end{document}